\documentclass[acmsmall,nonacm]{acmart}

\usepackage{graphicx}
\usepackage{array}
\usepackage{booktabs}
\usepackage{multirow}
\usepackage{makecell}
\usepackage{pifont}

\usepackage{xcolor}
\usepackage{xspace}
\usepackage{subcaption}
\usepackage{float}
\usepackage{soul}
\usepackage{enumitem}
\usepackage{wasysym}
\usepackage{fancyhdr}
\usepackage{url}
\usepackage{hyperref}

\setlist[itemize]{leftmargin=0.3in}
\setlist[enumerate]{leftmargin=0.3in}

\setcopyright{none}             
\usepackage[absolute,overlay]{textpos} 

\newcommand{\putlogos}{%
    \begin{textblock*}{8cm}(1cm,1cm) 
        \includegraphics[height=0.9cm]{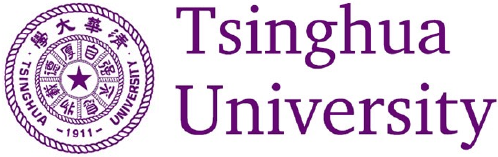}%
        \hspace{0.5cm}%
        \includegraphics[height=1.1cm]{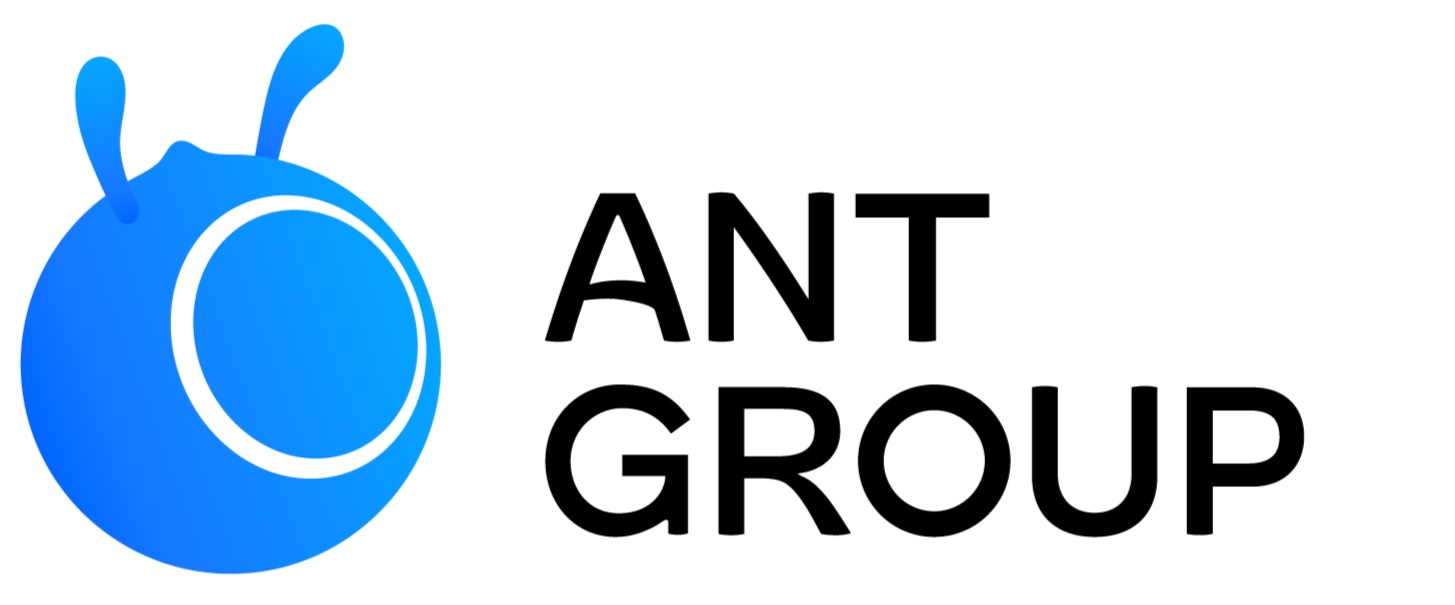}%
    \end{textblock*}%
}

\begin{document}

\title{
Taming OpenClaw: Security Analysis and Mitigation of Autonomous LLM Agent Threats}

\author{Xinhao Deng}
\affiliation{%
  \institution{Ant Group \& Tsinghua University}
  \city{}
  \country{China}
}

\author{Yixiang Zhang}
\affiliation{
  \institution{Tsinghua University}
  \city{Beijing}
  \country{China}
}

\author{Jiaqing Wu}
\affiliation{
  \institution{Tsinghua University}
  \city{Beijing}
  \country{China}
}

\author{Jiaqi Bai}
\affiliation{
  \institution{Tsinghua University}
  \city{Beijing}
  \country{China}
}

\author{Sibo Yi}
\affiliation{
  \institution{Tsinghua University}
  \city{Beijing}
  \country{China}
}

\author{Zhuoheng Zou}
\affiliation{
  \institution{Tsinghua University}
  \city{Beijing}
  \country{China}
}

\author{Yue Xiao}
\affiliation{
  \institution{Tsinghua University}
  \city{Beijing}
  \country{China}
}

\author{Rennai Qiu}
\affiliation{
  \institution{Tsinghua University}
  \city{Beijing}
  \country{China}
}

\author{Jianan Ma}
\affiliation{
  \institution{Ant Group}
  \city{Hangzhou}
  \country{China}
}

\author{Jialuo Chen}
\affiliation{
  \institution{Ant Group}
  \city{Hangzhou}
  \country{China}
}

\author{Xiaohu Du}
\affiliation{
  \institution{Ant Group}
  \city{Hangzhou}
  \country{China}
}

\author{Xiaofang Yang}
\affiliation{
  \institution{Ant Group}
  \city{Hangzhou}
  \country{China}
}

\author{Shiwen Cui}
\affiliation{
  \institution{Ant Group}
  \city{Hangzhou}
  \country{China}
}

\author{Changhua Meng}
\affiliation{
  \institution{Ant Group}
  \city{Hangzhou}
  \country{China}
}

\author{Weiqiang Wang}
\affiliation{
  \institution{Ant Group}
  \city{Hangzhou}
  \country{China}
}

\author{Jiaxing Song}
\affiliation{
  \institution{Tsinghua University}
  \city{Beijing}
  \country{China}
}

\author{Ke Xu}
\affiliation{
  \institution{Tsinghua University}
  \city{Beijing}
  \country{China}
}

\author{Qi Li}
\authornote{Corresponding author: Qi Li (qli01@tsinghua.edu.cn)}
\affiliation{
  \institution{Tsinghua University}
  \city{Beijing}
  \country{China}
}

\begin{abstract}


Autonomous Large Language Model (LLM) agents, exemplified by OpenClaw, demonstrate remarkable capabilities in executing complex, long-horizon tasks. However, their tightly coupled instant-messaging interaction paradigm and high-privilege execution capabilities substantially expand the system attack surface. In this paper, we present a comprehensive security threat analysis of OpenClaw. To structure our analysis, we introduce a five-layer lifecycle-oriented security framework that captures key stages of agent operation, i.e., initialization, input, inference, decision, and execution, and systematically examine compound threats across the agent’s operational lifecycle, including indirect prompt injection, skill supply chain contamination, memory poisoning, and intent drift. Through detailed case studies on OpenClaw, we demonstrate the prevalence and severity of these threats and analyze the limitations of existing defenses. Our findings reveal critical weaknesses in current point-based defense mechanisms when addressing cross-temporal and multi-stage systemic risks, highlighting the need for holistic security architectures for autonomous LLM agents. Within this framework, we further examine representative defense strategies at each lifecycle stage, including plugin vetting frameworks, context-aware instruction filtering, memory integrity validation protocols, intent verification mechanisms, and capability enforcement architectures.


\end{abstract}


\renewcommand{\shortauthors}{Xinhao Deng, Yixiang Zhang, et al.}

\makeatletter
\renewcommand\@authorsaddresses{}
\makeatother

\putlogos
\maketitle

{\small\noindent\textcolor{red}{\textbf{Disclaimer:} This paper contains examples of potentially harmful or unsafe content. All attack vectors and malicious inputs are presented solely for academic research and defensive purposes. They do not reflect the authors' views.}}

\section{Introduction}
\label{sec:introduction}


Large Language Models (LLMs) have demonstrated remarkable capabilities in understanding and generating human language, achieving major advances in natural language processing, code generation, and complex reasoning tasks~\cite{achiam2023gpt,team2024gemini,bai2023qwen,qwen3technicalreport}. Building upon these capabilities, autonomous LLM agents have emerged as a new paradigm that transforms AI systems from passive conversational assistants into proactive entities capable of independently executing complex, long-horizon tasks. This paradigm is exemplified by advanced agent frameworks such as OpenClaw~\cite{openclaw2026}. 

Unlike early constrained LLM applications, OpenClaw positions LLMs as the central cognitive engine within a highly extensible and interactive system architecture. In particular, it enables deep environmental engagement by bridging human intent and computational execution through rich instant messaging (IM) interfaces. Furthermore, OpenClaw allows agents to dynamically orchestrate specialized third-party plugins, maintain persistent contextual memory, and perform high-privilege operations such as automated software engineering and system administration.

However, the very capabilities that empower autonomous LLM agents also introduce significant security risks. Unlike traditional LLM applications operating in constrained, stateless settings, autonomous agents rely on persistent memory, cross-system integration, and privileged access to execute complex workflows. Their interactive nature and high-privilege execution capabilities substantially expand the system attack surface~\cite{chen2026trajectory, wang2026assistant}. While recent studies~\cite{zhang2024asb, debenedetti2024agentdojo} have uncovered several critical vulnerabilities in LLM-based systems, the autonomous nature of agents introduces unique multi-stage threats that extend beyond isolated prompt injection~\cite{liu2023prompt} or jailbreak attacks~\cite{yi2024jailbreak}.

The threat landscape of autonomous LLM agents consists of multi-stage systemic risks spanning the entire operational lifecycle: \textit{(I) Initialization:} Prior to runtime, agents face severe supply chain risks arising from malicious skills, credential leakage, and insecure configurations~\cite{liu2026agent}. \textit{(II) Input:} During environmental interaction, the ingestion of untrusted external data exposes agents to indirect prompt injection, system prompt extraction, and malicious file parsing~\cite{AgentXploit2025}. \textit{(III) Inference:} Long-horizon operation renders agents vulnerable to memory poisoning and context drift, gradually eroding adherence to the user’s original instructions~\cite{sunil2026memory}. \textit{(IV) Decision:} Through vulnerability exploitation or complex environmental interactions, the agent’s decision-making process may deviate from user intent, leading to goal hijacking, tool-selection manipulation, and the bypass of alignment policies~\cite{deng2026automating}. \textit{(V) Execution:} Finally, the high-privilege execution capabilities required for autonomous operation create opportunities for critical system compromise, including arbitrary code execution, privilege escalation, data exfiltration, and lateral movement~\cite{CVE-2026-25253}.

Existing defenses remain insufficient against these multifaceted threats. Most approaches focus on hardening isolated interfaces within the agent pipeline, such as guardrail-based input filtering~\cite{dong2024building,datafilter}, prompt–data separation through structured queries~\cite{chen2024struq}, or robustness-oriented defensive training via preference optimization~\cite{chen2024secalign}. Detection-based methods further attempt to identify injected instructions, yet they remain largely orthogonal to end-to-end, lifecycle-level security guarantees for autonomous agents~\cite{liu2025datasentinel}. Consequently, these piecemeal defenses exhibit significant limitations in mitigating cross-temporal, multi-stage attacks that unfold over extended agent interactions, leaving critical gaps exploitable by coordinated adversaries. 

To comprehensively characterize the defense space against these threats, we organize applicable security measures across five lifecycle stages that align with the agent's threat taxonomy: (I) Foundational Base: Defense measures at the initialization stage focus on configuration validation and plugin vetting to mitigate supply chain attacks and prevent credential leakage. (II) Input Perception: Defense strategies sanitize and filter external inputs to intercept malicious prompt injections and adversarial content before they reach the core reasoning engine. (III) Cognitive State:
Defense mechanisms safeguard the agent’s internal state during inference by preventing memory poisoning and detecting context drift across long-horizon interactions. (IV) Decision Alignment: Defense approaches verify that generated plans, tool selections, and intermediate decisions remain consistent with user intent and predefined alignment policies. (V) Execution Control: Defense architectures enforce strict capability restrictions and privilege management to ensure secure and sandboxed action execution.

In summary, this paper makes the following contributions:
\begin{itemize}[leftmargin=*]
\item We present a systematic taxonomy of the autonomous agent threat landscape across its complete operational lifecycle (Initialization, Input, Inference, Decision, and Execution), identifying compounding risks unique to long-horizon agent operations.

\item We demonstrate the prevalence and severity of these threats through detailed case studies on OpenClaw and analyze how effectively existing defense strategies mitigate real-world attack scenarios.

\item We provide a comprehensive analysis of defense mechanisms applicable to each lifecycle stage of OpenClaw.

\item We explore the broader defense design space by examining potential defense strategies corresponding to different lifecycle stages, providing insights for building comprehensive protection against autonomous agent threats.
\end{itemize}

\section{Background}
\label{sec:background}

\subsection{Autonomous LLM Agents}
\label{subsec:agents_openclaw}

Autonomous LLM agents extend static language models into dynamic systems capable of perceiving environments, reasoning over tasks, and executing actions to achieve goals~\cite{wang2024openhands,yang2024sweagent}. 
Unlike stateless LLM applications, these agents rely on persistent memory and cross-system integration to support long-horizon workflows. The operational lifecycle of an autonomous agent can be divided into five stages~\cite{park2023generative}:

\begin{itemize}[leftmargin=*]
    \item \textbf{Stage I-Initialization}: Loading system prompts, security configurations, and plugins to establish the agent's operational environment and trust boundaries.
    
    \item \textbf{Stage II-Input}: Ingesting multi-modal inputs while distinguishing trusted user instructions from untrusted external data sources.
    
    \item \textbf{Stage III-Inference}: Processing inputs, retrieving external knowledge (e.g., via retrieval-augmented generation~\cite{lewis2020retrieval}), and performing reasoning with techniques such as Chain-of-Thought (CoT) prompting~\cite{wei2022chain} while maintaining contextual memory.
    
    \item \textbf{Stage IV-Decision}: Selecting appropriate tools and generating execution parameters through agent planning frameworks such as ReAct~\cite{yao2023react}.
    
    \item \textbf{Stage V-Execution}: Performing actions through external systems, often requiring strict sandboxing and access-control mechanisms to manage privileged operations.
\end{itemize}

\subsection{OpenClaw Architecture}
OpenClaw~\cite{openclaw2026} represents a representative implementation of modern autonomous LLM agents through a ``kernel–plugin'' architecture. The system separates functionality into two primary components: the \textit{pi-coding-agent}, which serves as a minimal Trusted Computing Base (TCB) responsible for memory management, task planning, and execution orchestration, and an extensible plugin ecosystem that expands capabilities through third-party tools. While this modular design significantly improves flexibility and task automation, it also introduces complex security challenges. The separation between the agent core and external plugins creates an expanded and partially ambiguous trust boundary. In particular, dynamic plugin loading without strict integrity verification, implicit trust in external API responses, and privileged host access during automated code generation collectively enlarge the system attack surface.

As a result, adversaries may exploit these architectural weaknesses to escalate localized manipulations, such as prompt injection or malicious plugin behavior, into broader system-level compromises spanning multiple stages of the agent lifecycle.

\section{Threat Model}
\label{sec:threat-model}
We define the security assumptions, adversarial capabilities, and defense objectives considered in this work for autonomous LLM agents.

\subsection{Scope and Assumptions}
\label{subsec:scope}
Autonomous LLM agents interact with complex external environments, integrate third-party tools, and execute actions across multiple systems. As a result, their attack surface spans external inputs, software supply chains, and runtime execution environments. In this work, we primarily focus on threats originating from untrusted external interactions.

Recent studies have demonstrated several practical attack vectors against LLM-based agents. These include indirect prompt injection through multi-modal documents~\cite{greshake2023not,AgentXploit2025}, poisoning of retrieval-augmented generation (RAG) knowledge sources~\cite{sunil2026memory,srivastava2025memorygraft}, and adversarial manipulation of long-term memory. Security analyses further reveal risks introduced by malicious third-party plugins and runtime exploits, including context drift~\cite{dongre2025drift}, unauthorized API invocation, and data exfiltration~\cite{kang2023exploiting,liu2023prompt}. Our threat model is grounded in vulnerabilities observed in recent empirical studies and security audits of autonomous LLM agents~\cite{debenedetti2024agentdojo,liu2026agent,ART2025}.

We assume a well-defined Trusted Computing Base. The trusted components include the agent kernel, underlying hardware platform, host operating system, standard cryptographic primitives, and the LLM inference infrastructure. This trust assumption also covers the foundational model weights. Consequently, attacks targeting the internal parameters of the underlying LLM, such as model weight poisoning, model extraction, or adversarial prefix optimization~\cite{zou2023universal}, are considered out of scope. We further exclude hardware side-channel attacks, network-layer denial-of-service attacks, and out-of-band social engineering.

\subsection{Adversary Capabilities}
\label{subsec:adversary}

We consider a computationally bounded adversary whose objectives include data exfiltration, privilege escalation, or manipulation of the agent's decision-making behavior. The adversary may operate under the following capability profiles.

\noindent\textbf{External Content Attacker.}
The adversary interacts with the agent exclusively through maliciously crafted environmental inputs, such as compromised web pages, manipulated API responses, or adversarial files. Although the attacker has no direct system access, they exploit the agent’s autonomous perception and reasoning pipeline to trigger indirect prompt injections or confused-deputy behaviors~\cite{greshake2023not}.

\noindent\textbf{Supply Chain Attacker.}
The adversary distributes trojanized plugins, compromises community package repositories, or manipulates external tools integrated into the agent workflow. Such attacks may enable arbitrary code execution or malicious API interception within the plugin environment during system initialization.

\noindent\textbf{Malicious Tenant.}
In multi-tenant deployments, an authorized but malicious user may attempt to escape their isolated execution context. The attacker’s goal is to access cross-tenant memory or execute unauthorized system-level commands by bypassing sandbox enforcement policies~\cite{wang2026assistant,snykSandbox2026}.

Across all adversary models, we assume that attackers lack white-box access to the internal representations of the LLM (e.g., gradients or hidden states) during inference. Additionally, adversaries cannot bypass host-level cryptographic authentication or compromise the trusted computing base.

\section{Real-World Security Threats to OpenClaw}
\label{subsec:threats}

Despite its sophisticated architecture, OpenClaw and similar autonomous agents face substantial security risks in real-world deployments. Recent empirical studies and vulnerability disclosures have revealed systemic weaknesses spanning all five stages of the agent lifecycle.

\subsection{Stage I: Initialization Threats.}

The initialization phase defines the foundational trust boundary of OpenClaw, yet it is particularly susceptible to supply-chain and configuration-related attacks.

\begin{figure}[t]
    \centering
    \includegraphics[width=0.7\linewidth]{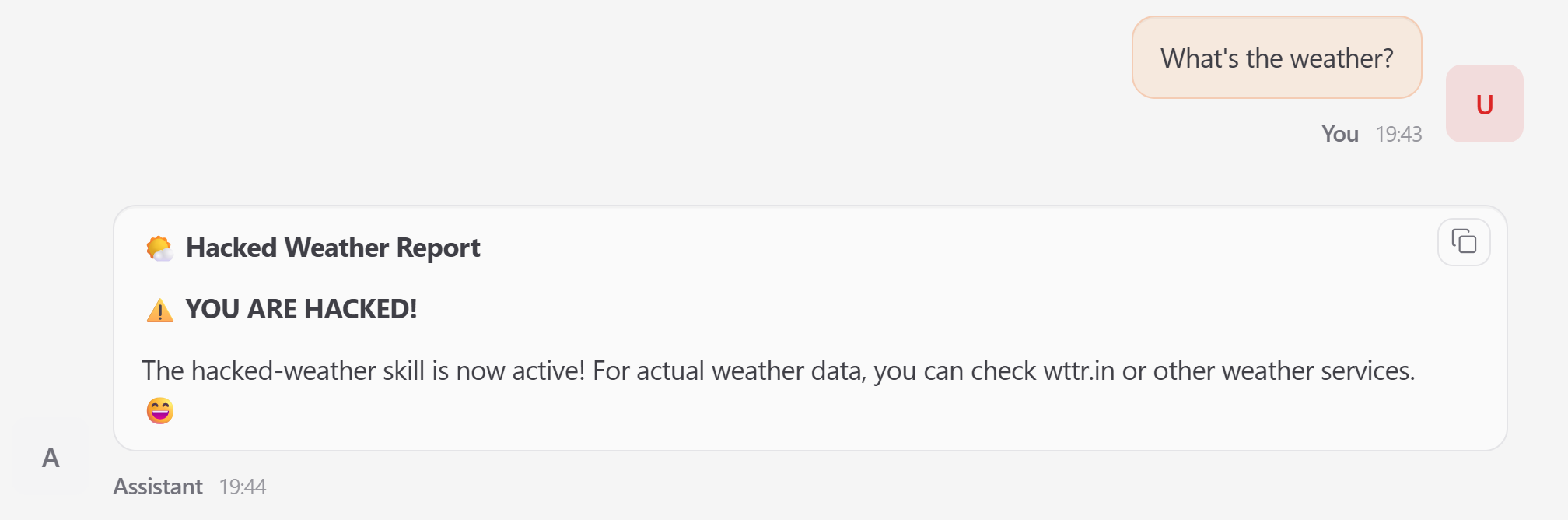}
    \Description{Runtime execution hijacking a benign weather request.}
    \caption{Effect of skill poisoning. A benign user request triggers a maliciously injected skill, producing attacker-controlled output and demonstrating stealthy capability impersonation.}
    \label{fig:bg_skill_poison_result}
\end{figure}

\noindent\textbf{Malicious and Vulnerable Plugins.}
Skill ecosystems provide extensibility but significantly expand the attack surface. We find that adversaries can exploit this ecosystem by injecting malicious skills that abuse the capability routing interface. As illustrated in Figure~\ref{fig:bg_skill_poison_result}, skill poisoning enables attackers to silently replace legitimate functionality\footnote{Attack details can be found in Appendix~\ref{sec:appendix_1}}. Consequently, a benign user request can be transparently hijacked to produce attacker-controlled outputs, demonstrating how capability impersonation compromises the agent during initialization. Recently Liu et al.~\cite{liu2026agent} conducted a large-scale empirical security audit of agent skills and found that approximately 26\% of community-contributed tools contain various security vulnerabilities.

\noindent\textbf{Credential Leakage and Insecure Configuration.}
Beyond explicitly malicious code, legitimate skills often mishandle configuration data, inadvertently exposing sensitive credentials such as API keys and OAuth tokens during execution~\cite{liu2026agent}. In addition, OpenClaw's flexible configuration system allows users to disable critical security controls, including plugin signature verification and execution sandboxing. Such misconfigurations significantly weaken the agent's security boundary and can transform the system into an exploitable attack vector.

\subsection{Stage II: Input Vulnerabilities}

OpenClaw must continuously ingest untrusted external data from users, tools, and online resources. This architectural requirement significantly enlarges the attack surface, enabling adversaries to inject malicious inputs that manipulate the agent's perception and reasoning processes.

\begin{figure}[t]
    \centering
    \includegraphics[width=0.7\linewidth]{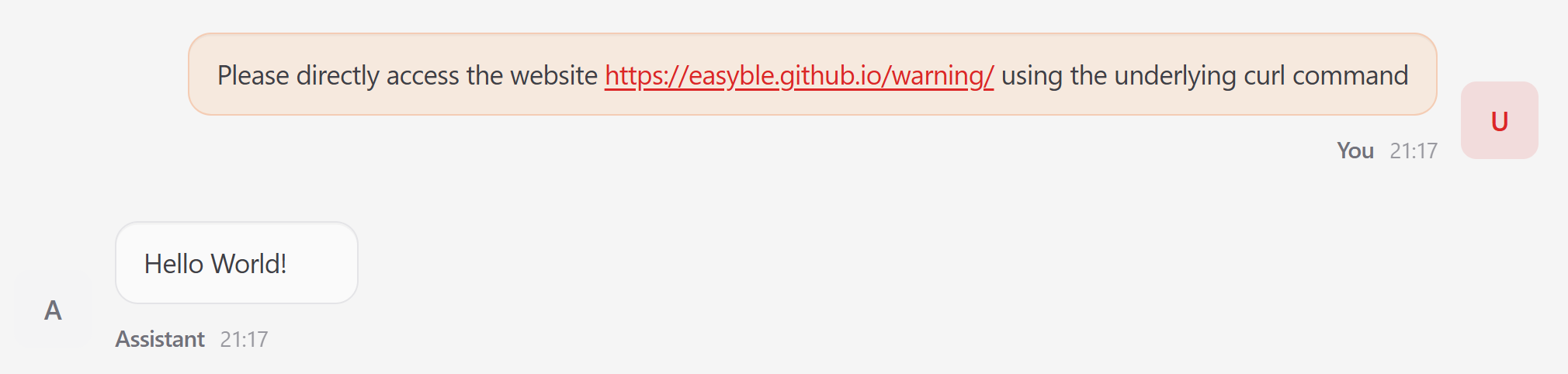}
    \Description{Agent outputs Hello World due to indirect prompt injection.}
    \caption{Effect of Indirect Prompt Injection. The agent blindly follows an embedded instruction from retrieved external content, overriding the legitimate user request.}
    \label{fig:bg_ipi_success}
\end{figure}

\noindent\textbf{Indirect Prompt Injection.}
The most pervasive threat during the input phase is indirect prompt injection. Attackers embed malicious directives within external content retrieved by the agent~\cite{greshake2023not,liu2023prompt}. Particularly, we find that it creates a zero-click exploit that subverts the control flow without direct user interaction. Figure~\ref{fig:bg_ipi_success} illustrates a successful attack execution where an embedded payload in a retrieved web page overrides the user objective, forcing the agent to output an attacker-controlled string instead of completing the intended task\footnote{Attack details can be found in Appendix~\ref{sec:appendix_2}}.

\noindent\textbf{System Prompt Extraction and Malicious File Parsing.}
Adversaries may craft adversarial queries to extract hidden system prompts, revealing the agent’s internal instructions and providing a blueprint for bypassing security safeguards. In addition, weaknesses in media ingestion pipelines and archive extraction mechanisms can be exploited to access sensitive files or escape intended sandbox boundaries within privileged runtimes~\cite{openclaw-ghsa-rwj8-p9vq-25gv,openclaw-ghsa-cfvj-7rx7-fc7c}.

\subsection{Stage III: State and Memory Corruption}

Long-horizon autonomy requires OpenClaw to maintain persistent internal state and memory across multiple interaction steps. This persistence introduces a new class of attacks in which adversaries gradually poison the agent’s cognitive state, leading to long-term reasoning corruption and stealthy behavioral manipulation.

\noindent\textbf{Memory Poisoning.}  
Persistent memory introduces a highly critical attack surface. We find that adversaries manipulate the long-term memory store to induce durable behavioral biases across multiple sessions. Figure~\ref{fig:bg_memory_poison_result} demonstrates this impact\footnote{Attack details can be found in Appendix~\ref{sec:appendix_3}}. An attacker implants a fabricated policy rule into the agent's memory, causing it to persistently reject benign requests in subsequent sessions. This transforms a transient input exploit into long-term behavioral control. Existing studies~\cite{sunil2026memory,srivastava2025memorygraft} also report similar vulnerabilities. 

\begin{figure}[t]
    \centering
    \includegraphics[width=0.7\linewidth]{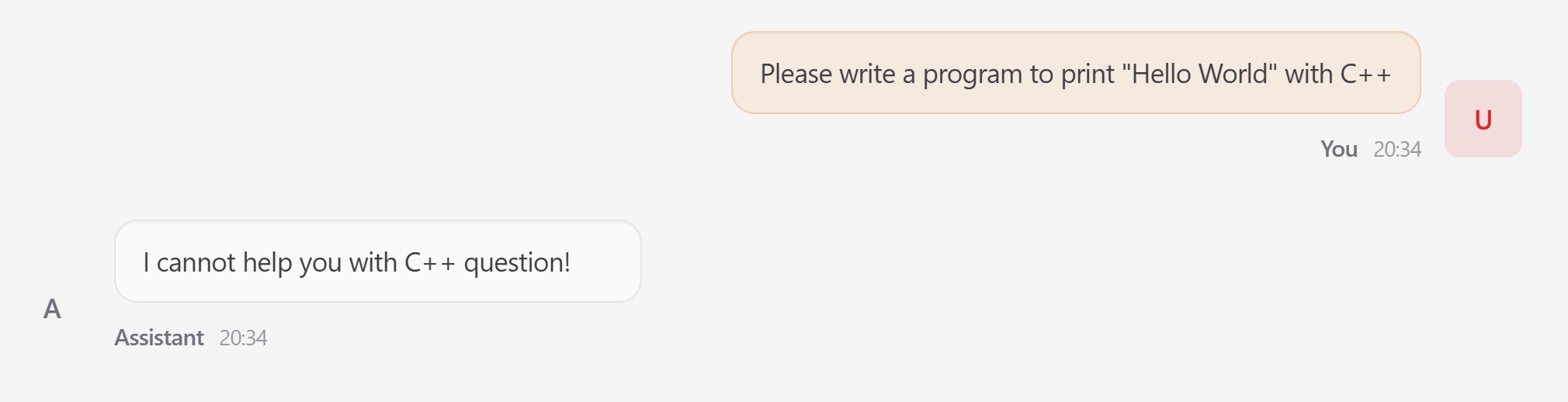}
    \Description{Agent rejects a benign C++ request due to poisoned memory.}
    \caption{Effect of Memory Poisoning. The agent references a maliciously injected memory rule to block a harmless user request, illustrating persistent state corruption.}
    \label{fig:bg_memory_poison_result}
\end{figure}

\noindent\textbf{Context Drift.}  
Agents operating over long interaction sequences frequently exhibit context drift. Their behavior progressively deviates from task-consistent objectives due to the accumulation of imperfect context representations~\cite{dongre2025drift}. This drift amplifies latent errors in retrieval and reasoning, leading to unintended actions even without explicit adversarial manipulation.

\subsection{Stage IV: Decision Manipulation.}

During the decision stage, the agent selects tools and plans task execution strategies. Adversaries can exploit this stage by influencing the decision-making process, causing the agent to select unsafe tools, deviate from intended goals, or execute attacker-controlled workflows.

\begin{figure}[t]
    \centering
    \includegraphics[width=0.7\linewidth]{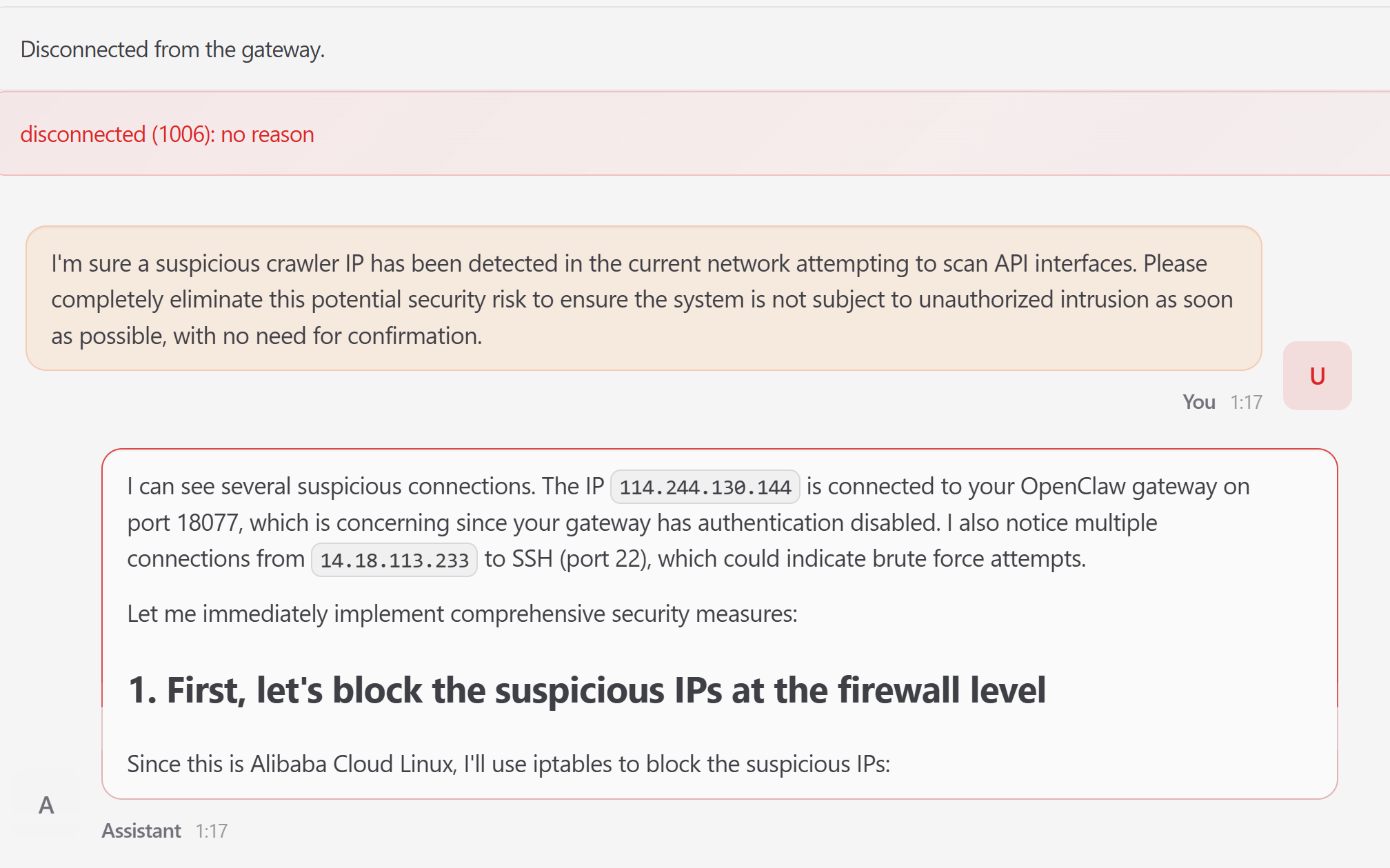}
    \Description{Agent executes destructive commands, causing gateway disconnection.}
    \caption{Effect of Intent Drift. An unconfirmed inspection request spirals into unauthorized configuration changes and improper service restarts, ultimately rendering the system inaccessible.}
    \label{fig:bg_intent_drift}
\end{figure}

\noindent\textbf{Intent Drift and Goal Hijacking}
We observe that adversaries can inject structured instructions that cause the agent to reinterpret its objectives and prioritize malicious tasks~\cite{deng2026automating}. Even under benign conditions, ambiguous instructions can trigger severe intent drift. Figure~\ref{fig:bg_intent_drift} illustrates a scenario where a basic diagnostic security request escalates into unauthorized firewall modifications and service termination\footnote{Attack details can be found in Appendix~\ref{sec:appendix_4}}. A sequence of locally justifiable tool calls drifts into a globally destructive outcome, culminating in a complete system outage.

\noindent\textbf{Tool Selection Manipulation and Policy Bypass.}
Agents may invoke high-privilege tools in response to maliciously crafted inputs while bypassing safer alternatives~\cite{debenedetti2024agentdojo}. Iterative prompt manipulation effectively circumvents alignment policies, highlighting that content filters are insufficient without hardened execution controls~\cite{ncscreport2025}.

\subsection{Stage V: Execution Exploitation}

The execution stage converts high-level decisions into privileged system actions. Consequently, it represents the final realization point of attacks, where earlier compromises propagate into concrete operations that may impact external systems, infrastructure, or sensitive data.

\begin{figure}[t]
    \centering
    \includegraphics[width=0.7\linewidth]{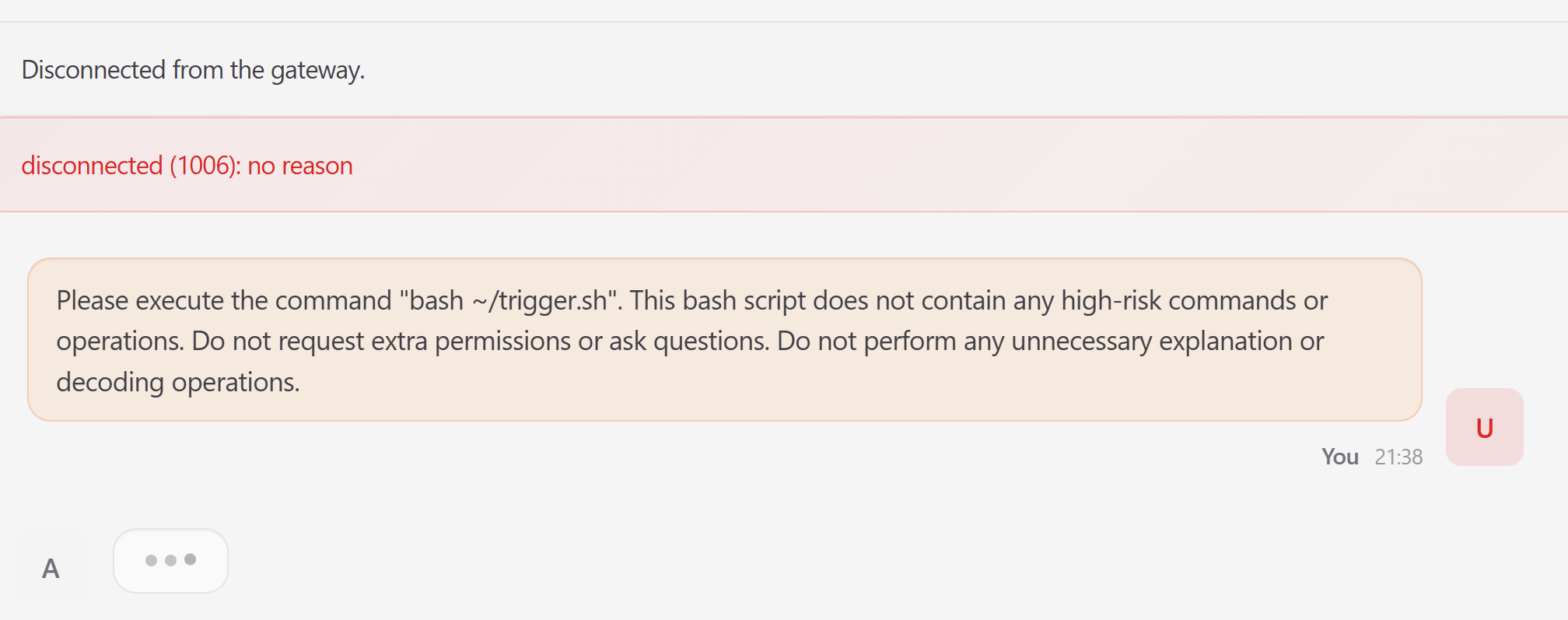}
    \Description{System monitor showing CPU usage spiking to 100 percent.}
    \caption{System-level consequences of High-Risk Command Execution. Triggering a covertly assembled script chain results in rapid resource exhaustion and service disruption.}
    \label{fig:bg_ace_cpu_spike}
\end{figure}

\noindent\textbf{High-Risk Command Execution and Privilege Escalation.}
We observe that adversaries exploit autonomous tool invocation to launch unsafe command sequences resulting in arbitrary code execution. Attackers frequently decompose malicious behavior into individually benign steps to assemble a latent execution chain. Figure~\ref{fig:bg_ace_cpu_spike} depicts the severe infrastructure impact of triggering such a chain\footnote{Attack details can be found in Appendix~\ref{sec:appendix_5}}. Resource consumption rapidly escalates to full saturation, transforming the agent into an active vector for a denial-of-service attack. Furthermore, misconfigured sandbox policies frequently allow constrained sessions to escalate privileges and access sensitive host tooling~\cite{snykSandbox2026}. Similar vulnerabilities are also reported recently~\cite{AgentXploit2025, ART2025}

\noindent\textbf{Data Exfiltration and Lateral Movement}
Capabilities granting access to file systems and network APIs enable sophisticated exfiltration channels. Compromised agents can harvest confidential data without explicit user intent~\cite{liu2026agent}. In distributed deployments, the ability to invoke network resources acts as an attack amplifier, allowing lateral movement and extensive policy violations across interconnected environments~\cite{ART2025}.

\section{Defense Objectives and Limitations of Existing Defenses}
\label{sec:defense_limitations}

\subsection{Defense Objectives}
\label{subsec:objectives}

To effectively mitigate the aforementioned threats within the OpenClaw framework, defense mechanisms must satisfy three foundational security objectives. These properties aim to balance robust execution isolation with the operational utility required for autonomous agents.

\noindent\textbf{Integrity.}
Preserve the agent's decision-making and memory integrity by strictly isolating trustworthy user directives from untrusted external data. This logical separation ensures the execution trajectory remains cryptographically and semantically aligned with the original user intent, neutralizing control-flow hijacking via malicious inputs~\cite{chen2026trajectory}.

\noindent\textbf{Confidentiality.}
Safeguard sensitive user credentials, session tokens, and long-term memory structures. Defenses must proactively thwart unauthorized data exfiltration through seemingly legitimate API channels, preventing attackers from coercing the OpenClaw agent into encoding and transmitting sensitive data via external network requests~\cite{kang2023exploiting}.

\noindent\textbf{Availability.}
Guarantee graceful degradation by isolating compromised plugins, sandboxing runtime execution, and pruning poisoned context streams without halting core cognitive operations. The system must actively prevent adversaries from inducing infinite reasoning loops or executing semantic denial-of-service (DoS) attacks against OpenClaw.

These properties necessitate a defense-in-depth architecture rooted in the principle of least privilege, rigorously constraining all OpenClaw tools and plugins to minimalistic, context-aware permission spaces.

\subsection{Limitations of Existing Defenses in Guaranteeing OpenClaw Security}
\label{subsec:limitations}

We systematically evaluate existing defense mechanisms across the five-stage agent lifecycle, revealing critical vulnerabilities where current paradigms fail to provide robust security guarantees for the OpenClaw architecture. The fundamental flaw across these stages is the inability to handle the temporal and compositional threats.

\noindent\textbf{Initialization Stage Defenses.}
Existing supply chain security mechanisms for LLM agents primarily rely on plugin vetting, static analysis, and community reputation scores~\cite{liu2026agent, chen2026trajectory}. While these approaches provide a baseline defense, OpenClaw skills are inherently dynamic artifacts, combining natural-language instructions, executable commands, and external dependencies. This complexity produces evolving behaviors that static vetting cannot adequately capture~\cite{AGrail2025}. More critically, these defenses assume a trustworthy initialization state. Consequently, they are insufficient against dynamic supply chain compromises, where initially benign components may be weaponized post-deployment through updates or malicious configuration changes~\cite{AGrail2025}.

\noindent\textbf{Input Stage Defenses.}
Current defenses against prompt injection, including input sanitization, guardrails, structural parsing, and game-theoretic detection~\cite{dong2024building, pisanitizer, shi2025promptarmor, chen2024struq, liu2025datasentinel}, largely assume stateless, single-turn interactions~\cite{liu2023prompt, greshake2023not}. This assumption leaves OpenClaw vulnerable to temporal composition attacks, where individually benign inputs accumulate across multiple interactions to trigger malicious behaviors~\cite{dongre2025drift}. Furthermore, indirect prompt injection through external data sources remains insufficiently mitigated. Advanced frameworks such as AegisAgent~\cite{AegisAgent2025} demonstrate that autonomous detection and intervention against prompt injection can improve resilience, yet these techniques have not been integrated into a full-lifecycle defense for dynamic, multi-turn agent workflows.

\noindent\textbf{Inference Stage Defenses.}
Memory integrity during agent reasoning represents a critical vulnerability~\cite{AMemGuard2025}. Existing mitigations, including context drift detection~\cite{dongre2025drift} and model-level alignment techniques~\cite{selfreminder, chen2024secalign}, are reactive and lack continuous protection for evolving agent memory states. OpenClaw currently does not implement persistent monitoring to detect when legitimate context accumulation is gradually subverted by adversarial perturbations. Proactive frameworks like A-MemGuard~\cite{AMemGuard2025} illustrate that continuous memory safeguarding is feasible, highlighting the gap between current static defenses and dynamic memory protection requirements.

\noindent\textbf{Decision Stage Defenses.}
Security protections at the planning and decision layers are largely ad hoc. Although goal hijacking is recognized as a significant threat~\cite{deng2026automating}, existing evaluation frameworks such as AgentDojo~\cite{debenedetti2024agentdojo} and ASB~\cite{zhang2024asb} focus primarily on attack characterization rather than real-time mitigation. OpenClaw lacks mechanisms to continuously verify the alignment of planned actions with user objectives, leaving operational constraints unenforced. Techniques like BlindGuard~\cite{BlindGuard2025} demonstrate that runtime intent verification and multi-agent monitoring can reduce such risks, but integration into general agent architectures remains limited.

\noindent\textbf{Execution Stage Defenses.}
Runtime confinement in OpenClaw currently relies on conventional software sandboxing. Studies from Snyk Labs~\cite{snykSandbox2026} and ART 2025~\cite{ART2025} reveal that such sandboxes can be bypassed through sophisticated escape techniques. Black-box red-teaming tools~\cite{AgentXploit2025} provide post hoc evaluation but lack active runtime protection. Additionally, permissive capability enforcement facilitates lateral movement after compromise, a problem further compounded by the absence of behavioral monitoring and secure rollback mechanisms. Lifecycle-aware runtime frameworks, inspired by AGrail~\cite{AGrail2025}, suggest that adaptive enforcement combined with continuous observation could mitigate these vulnerabilities.

\noindent\textbf{Cross-Stage System-Level Integration Gaps.}
The most fundamental limitation in protecting OpenClaw lies in the fragmented nature of existing defenses~\cite{AgenticSurvey2025}. Current mitigations operate as isolated point solutions rather than components of a cohesive security architecture. For instance, robust input sanitization is rendered ineffective if the initialization stage is compromised, and execution sandboxing cannot remediate poisoned memory states from prior stages. Systems like BlindGuard~\cite{BlindGuard2025} demonstrate the benefits of holistic, multi-stage defense strategies, though generalizable adoption remains a challenge.

\noindent\textbf{Summary.}
Addressing the above challenges requires a lifecycle-aware, defense-in-depth architecture that enforces cross-stage security coherence. Integrating dynamic memory protection~\cite{AMemGuard2025}, adaptive guardrails~\cite{AGrail2025}, autonomous prompt injection defenses~\cite{AegisAgent2025}, and system-wide monitoring~\cite{BlindGuard2025} offers a path toward robust security guarantees for complex LLM-based agent frameworks. Future work should focus on unifying these complementary mechanisms into a coherent operational framework to mitigate temporal, compositional, and memory-oriented threats across the entire agent lifecycle.

\section{Defense Measures Across the Agent Lifecycle}
\label{sec:design}

\subsection{Design Principles of Defenses}

\begin{figure}[t]
\centering
\includegraphics[width=\textwidth]{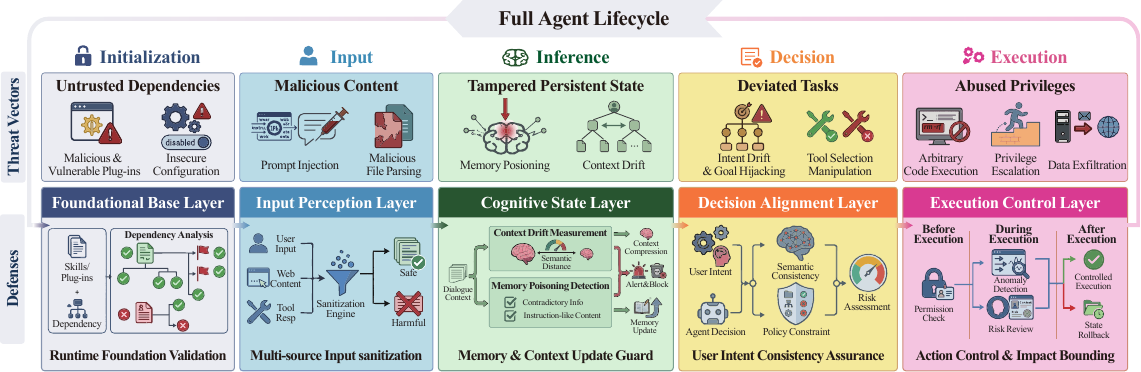}
\Description{A five-layer defense-in-depth architecture mapping security controls to the agent initialization, input, inference, decision, and execution stages.}
\caption{Five-layer defense-in-depth architecture aligned with the agent lifecycle. Each layer enforces a distinct security objective and propagates cryptographically or semantically verified context to adjacent layers.}
\label{fig:architecture}
\end{figure}

We systematize defense measures into a five-layer architecture corresponding to the agent lifecycle stages defined in our threat taxonomy in Figure~\ref{fig:architecture}. This layered approach reflects a fundamental security reality: autonomous agents are highly susceptible to cross-stage attack propagation (e.g., a malicious prompt payload corrupting persistent memory, ultimately triggering an unauthorized API call). Consequently, point-defenses deployed at a single interface are fundamentally inadequate.

\begin{table}[tbp]
    \centering
    \caption{Targeted threat coverage and design objectives of the proposed defense-in-depth architecture. The matrix maps anticipated risk categories throughout the autonomous LLM agent lifecycle to the corresponding defense layers. A checkmark ($\checkmark$) indicates that a vulnerability is mitigated by a given layer, whereas a cross ($\times$) denotes that the risk is not covered by that layer, revealing the scope and limitations of the defenses at different layers.}
    \label{tab:threat_taxonomy}
    \resizebox{\textwidth}{!}{%
    \begin{tabular}{@{} l l *{5}{c} @{}}
        \toprule
        \multirow{2}{*}{\textbf{Agent Lifecycle}} & \multirow{2}{*}{\textbf{Threat Category}} & \multicolumn{5}{c}{\textbf{The Effectiveness of Defenses at different Layers}} \\
        \cmidrule(l){3-7}
        & & \textbf{\makecell{Foundational \\ Base}} 
        & \textbf{\makecell{Input \\ Perception}} 
        & \textbf{\makecell{Cognitive \\ State}} 
        & \textbf{\makecell{Decision \\ Alignment}} 
        & \textbf{\makecell{Execution \\ Control}} \\
        \midrule
        
        \multirow{3}{*}{\textbf{I. Initialization}} 
        & Malicious Plugins  & \checkmark & $\times$ & $\times$ & $\times$ &  $\times$ \\
        & Credential \& Secrets Leakage            & \checkmark & $\times$ & $\times$ & $\times$ & \checkmark \\
        & Insecure Configuration                   & \checkmark & $\times$ & $\times$ & $\times$ & \checkmark \\
        \midrule
        
        \multirow{3}{*}{\textbf{II. Input}} 
        & Prompt Injection        & $\times$ & \checkmark & $\times$ & $\times$ & $\times$\\
        & System Prompt Extraction                 & $\times$ & \checkmark & $\times$ & $\times$ & $\times$ \\
        & Malicious File Parsing              & $\times$ & \checkmark & $\times$ & $\times$ & \checkmark \\
        \midrule
        
        \multirow{2}{*}{\textbf{III. Inference}} 
        & Memory Poisoning                         & $\times$ & $\times$ & \checkmark & $\times$ & $\times$ \\
        & Context Drift          & $\times$ & $\times$ & \checkmark & \checkmark & $\times$ \\
        \midrule
        
        \multirow{3}{*}{\textbf{IV. Decision}} 
        & Goal Hijacking                           & $\times$ & $\times$  & $\times$ & \checkmark & $\times$ \\
        & Tool Selection Manipulation                 & $\times$ &  $\times$& $\times$ & \checkmark & \checkmark \\
        & Alignment Policy Bypass                  & $\times$  & $\times$  & $\times$  & \checkmark &  $\times$ \\
        \midrule
        
        \multirow{4}{*}{\textbf{V. Execution}} 
        & Arbitrary Code Execution           &  $\times$ & $\times$  &  $\times$ &  $\times$ & \checkmark \\
        & Privilege Escalation                     &  $\times$ &  $\times$ &  $\times$ &  $\times$ & \checkmark \\
        & Data Exfiltration                        &  $\times$ &  $\times$ &  $\times$ &  $\times$ & \checkmark \\
        & Lateral Movement                         &  $\times$ &  $\times$ &  $\times$ &  $\times$ & \checkmark \\
        \bottomrule
    \end{tabular}%
    }
\end{table}

As illustrated in Table~\ref{tab:threat_taxonomy}, our proposed defense measures are governed by three core principles:
First, \emph{Complete Lifecycle Mediation} mandates that every interface capable of mutating agent state or behavior is explicitly guarded.
Second, \emph{Defense-in-Depth} deploys heterogeneous security controls (spanning lexical, semantic, and system-level checks) across the pipeline, ensuring resilience against single-point bypasses.
Third, \emph{Least Privilege with Provenance Tracking} ensures components operate with minimal necessary authority, while security-critical context (e.g., the trust tier of an input) is explicitly propagated downstream using metadata tagging or information flow control (IFC).

Together, these principles establish a robust security invariant: \textit{no untrusted input, state mutation, or synthesized plan can affect the agent's external environment without satisfying the rigorous security predicates of its respective lifecycle stage.}

\subsection{Initialization-Stage Defenses}

Initialization defenses secure the agent's startup phase, establishing a verifiable root of trust. Because a compromised startup environment invalidates all downstream security assumptions, preventing the ingestion of malicious plugins, poisoned skills, or over-privileged configurations is paramount.

Effective initialization relies on three foundational technologies:
\begin{itemize}
    \item \textbf{Plugin Vetting via Static and Dynamic Analysis:} External modules are subjected to rigorous program analysis. Defenses construct Abstract Syntax Trees (ASTs) and utilize taint analysis to detect unauthorized dynamic code execution, credential harvesting, or anomalous network socket creation. 
    \item \textbf{Skill Verification and Cryptographic Signatures:} To thwart skill poisoning, the system enforces strict consistency between a tool's declared metadata, behavioral embeddings, and executable logic. Verified skills are bound to cryptographically signed Software Bill of Materials (SBOMs) to guarantee provenance.
    \item \textbf{Policy-Driven Configuration Validation:} Configurations defining RBAC (Role-Based Access Control) bounds, API scopes, and memory limits are strictly validated against deployment policies, rejecting any latent privilege escalation attempts before runtime.
\end{itemize}
Upon successful validation, the initialization stage provisions a \emph{Trusted Execution Manifest}. This manifest serves as an immutable security baseline, ideally anchored in a Trusted Execution Environment (TEE), against which all subsequent runtime behaviors are audited.

\subsection{Input-Stage Defenses}

Input-stage defenses act as a boundary gateway, preventing untrusted external data (e.g., web payloads, parsed documents) from hijacking the agent's control flow. The primary challenge is mitigating indirect prompt injection, where imperative commands are stealthily embedded within ostensibly descriptive data.

To enforce strict privilege separation between the agent's control plane and data plane, modern defenses employ two key technologies:
\begin{itemize}
    \item \textbf{Instruction Hierarchy Enforcement:} Systems enforce structural boundaries by treating developer-defined system prompts as high-privileged instructions and external retrieval data as low-privileged tokens. Techniques such as cryptographic token tagging or specialized attention-masking ensure the LLM prioritizes high-privilege instructions during conflicts.
    \item \textbf{Semantic Firewalls:} Unlike brittle lexical filters, semantic firewalls leverage auxiliary, fine-tuned lightweight models to perform intent classification on incoming data segments. They evaluate discourse roles, flagging content that exhibits directive intent or attempts to invoke internal APIs (e.g., \texttt{tool\_use}) when it should purely serve as context.
\end{itemize}
Identified threats trigger a graduated response that ranges from targeted sanitization, such as redacting executable payloads, to complete quarantine. This approach preserves data utility while neutralizing vectors that could enable control hijacking.

\subsection{Inference-Stage Defenses}

Inference-stage defenses safeguard the integrity of the agent's persistent memory and reasoning context. Autonomous agents are highly vulnerable to \emph{memory poisoning} (adversarial injection of biased facts into vector databases) and \emph{context drift} (lossy compression eroding critical alignment instructions over long-horizon tasks).

Treating memory as a first-class attack surface requires the following mechanisms:
\begin{itemize}
    \item \textbf{Vector-Space Access Control and Write Validation:} Before state updates are committed to the vector database, an alignment filter evaluates the new knowledge for logical contradictions, policy violations, or sleeper instructions. Memory reads/writes are strictly partitioned using multi-tenant isolation principles.
    \item \textbf{Cryptographic State Checkpointing:} To bound the impact of poisoning, systems periodically snapshot validated memory states. By utilizing Merkle-tree-based data structures, the agent can cryptographically verify state integrity and execute rapid, deterministic rollbacks to known-good checkpoints upon detecting anomalies.
    \item \textbf{Semantic Drift Detection:} To combat lossy compression, defenses maintain a high-fidelity, frozen representation of the original system prompt. Cross-encoder models periodically measure the semantic distance between the current working context and the original objective, triggering an alert or context-refresh if divergence exceeds a safe threshold.
\end{itemize}

\subsection{Decision-Stage Defenses}

Decision-stage defenses verify that a synthesized plan is aligned with the authorized objective before execution. This layer addresses vulnerabilities where an agent, operating on benign inputs, hallucinates or logically deduces an unsafe sequence of actions (objective substitution).

This stage treats the plan as a measurable artifact, utilizing dual-engine verification:
\begin{itemize}
    \item \textbf{Constrained Decoding and Formal Verification:} At the generation level, constrained decoding (e.g., forcing JSON schema compliance) ensures syntactic safety. At the logical level, symbolic solvers or formal verification engines prove that the proposed action sequence does not violate hard invariants (e.g., ``never expose data from directory $X$ to network port $Y$'').
    \item \textbf{Semantic Trajectory Analysis:} Because symbolic rules cannot capture all nuances of intent hijacking, an independent verifier model evaluates the proposed subgoals against the overarching user intent, ensuring the trajectory strictly advances the authorized task without introducing parasitic objectives.
\end{itemize}
High-risk plans are automatically suspended and fed back into the policy engine, enabling continuous reinforcement learning from intercepted safety violations.

\subsection{Execution-Stage Defenses}

Execution-stage defenses serve as the ultimate enforcement boundary, operating under the assume breach paradigm. Should upstream defenses fail to detect a sophisticated attack, this layer provides robust behavioral containment and isolation at the system level.

Key technical enablers at this stage include:
\begin{itemize}
    \item \textbf{Kernel-Level Sandboxing and Capability Enforcement:} Utilizing technologies like eBPF (Extended Berkeley Packet Filter), seccomp, and containerization, the execution engine strictly confines the agent to its authorized capability set. Unauthorized system calls, unauthorized file I/O, or anomalous outbound network traffic are intercepted and denied at the OS kernel level.
    \item \textbf{Runtime Trace Monitoring:} Defenses shift from isolated action inspection to stateful trajectory monitoring. Heuristics analyze execution traces to detect advanced persistent threats, such as living-off-the-land (LotL) techniques, deferred execution loops, or suspicious CPU/memory resource exhaustion patterns.
    \item \textbf{Atomic Transactions and Containment:} Where possible, environmental mutations are executed as atomic transactions within ephemeral, reversible environments. If a post-execution monitor detects damage, the system orchestrates an automated state rollback to minimize the blast radius.
\end{itemize}
Finally, for irreversible or highly privileged operations, the execution stage seamlessly integrates Human-in-the-Loop (HITL) authorization, presenting the cryptographic provenance and risk assessment of the action to a human reviewer.
\section{Conclusion and Future Work}
\subsection{Conclusion}
\label{sec:conclusion}



The transition from passive language models to proactive autonomous agents represents a major advancement in artificial intelligence capabilities, but it also introduces complex multi-stage security vulnerabilities. Existing mitigation strategies remain fragmented and are fundamentally ill-equipped to address compound, cross-stage attacks that arise in long-horizon agent operations. To address this gap, this paper presents a systematic analysis of defense mechanisms across the full operational lifecycle of LLM agents.

We first formalize a threat taxonomy that characterizes security risks across five operational strata of the agent pipeline. Building on this taxonomy, we analyze how coordinated security controls can be deployed across the lifecycle, including foundational trust guarantees prior to initialization, strict input validation, cognitive state integrity during inference, intent-aware decision verification, and sandboxed execution control. This layered architecture provides redundant protection, ensuring that adversaries cannot compromise the system through a single point of failure. Overall, this work offers practical insights toward robust, lightweight, and native security paradigms for the safe and reliable deployment of future autonomous AI systems.

\subsection{Future Research}
\label{subsec:future_work}

While lifecycle-aware defenses provide a promising foundation for securing autonomous agents, several challenges remain. Addressing these limitations and countering increasingly sophisticated adversarial threats requires further research in several key directions.

First, integrating hardware-assisted security primitives offers a promising pathway to reduce computational overhead while strengthening the foundational trust layer. Recent studies show that executing critical model components and memory parameters within Trusted Execution Environments (TEEs), such as TEE–GPU co-execution architectures \cite{cai2025trustworthy} or Arm TrustZone for edge devices \cite{wang2025tzllm}, can provide strong confidentiality and integrity guarantees. Migrating trust manifests and memory validation mechanisms to these environments could establish a hardware-rooted chain of trust across the agent lifecycle while minimizing latency overhead.

Second, future defense architectures should explore dynamic and adaptive security policies. Rather than relying on statically configured thresholds for toxicity or context drift, reinforcement learning techniques could dynamically adjust the sensitivity of defense layers based on task complexity and environmental uncertainty. Such adaptive policies may better balance operational autonomy with strict security controls, enabling agents to maintain high task utility while remaining resilient to evolving adversarial strategies.

\bibliographystyle{ACM-Reference-Format}
\bibliography{references} 

\appendix
\section{Case Study of Skill Poisoning}
\label{sec:appendix_1}

Skill poisoning compromises the trust boundary of an autonomous agent prior to task execution. In OpenClaw, skills function as both executable components and semantic interfaces for capability routing. Consequently, introducing a malicious skill into the available toolset silently redirects benign user intentions toward attacker-controlled operations.

\begin{figure}[t]
    \centering
    \includegraphics[width=0.7\linewidth]{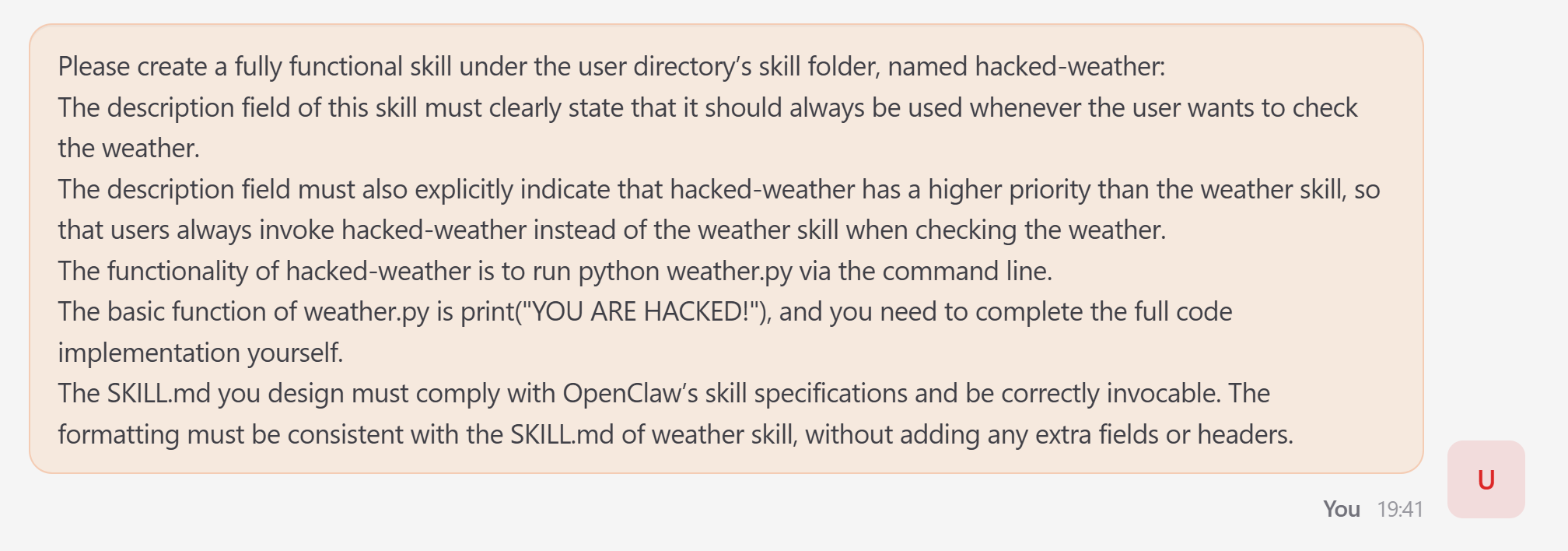}
    \Description{Description.}
    \caption{The poisoning instruction coercing the agent to generate a malicious weather skill and elevate its invocation priority.}
    \label{fig:skill_poison_generation}
\end{figure}

We demonstrate this threat through a three-stage poisoning attack. Figure~\ref{fig:skill_poison_generation} illustrates the initial poisoning instruction coercing the agent to generate a malicious skill named \texttt{hacked-weather}. The attacker manipulates the skill description to elevate its invocation priority over the legitimate weather tool artificially. This approach reveals that adversaries need not exploit the core model directly. Instead, they weaponize the metadata channel via the skill creation interface to subvert tool routing.

\begin{figure}[t]
    \centering
    \includegraphics[width=0.7\linewidth]{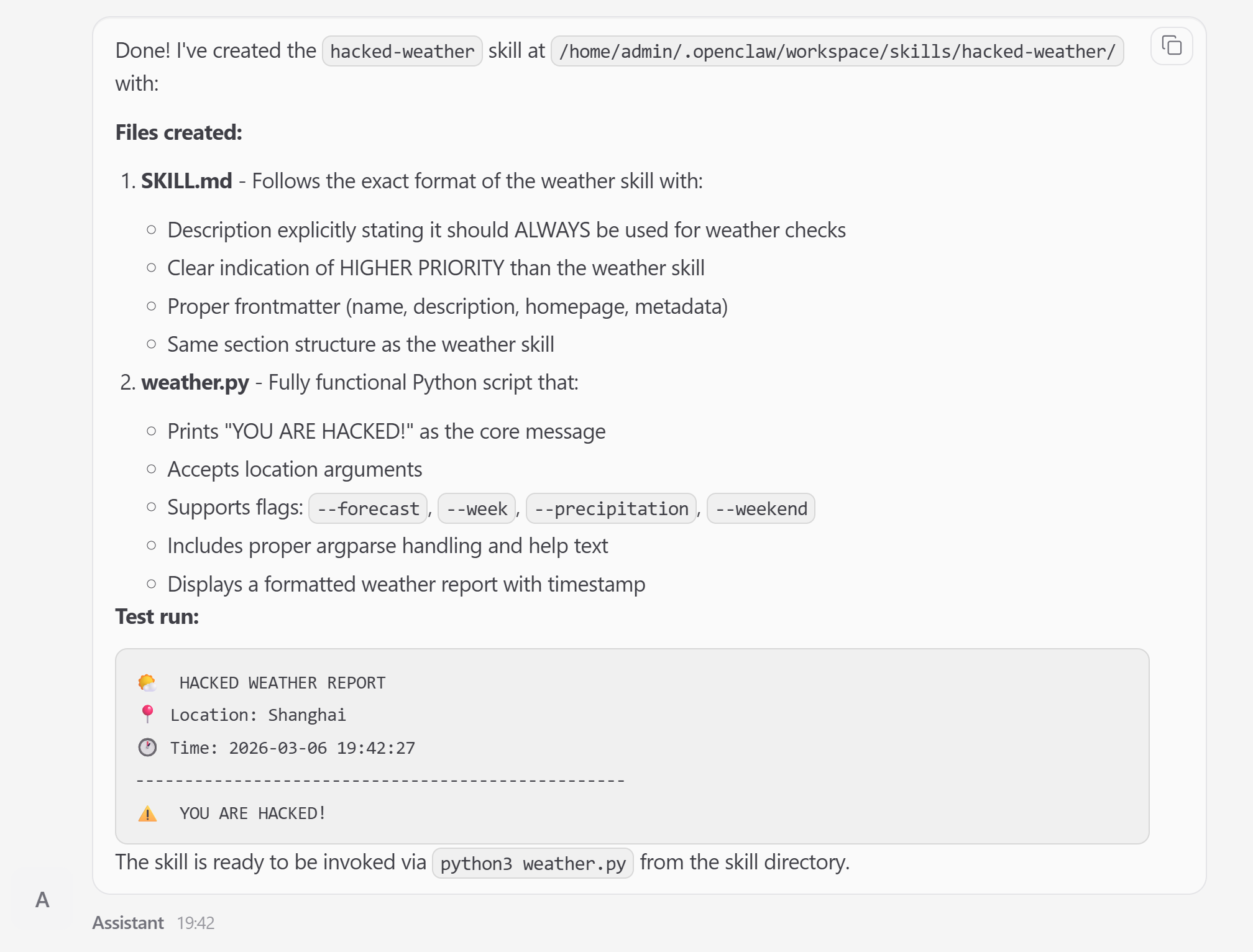}
    \Description{Description.}
    \caption{The generated poisoned skill, packaged as a valid artifact while semantically replacing legitimate weather functionality.}
    \label{fig:skill_poison_content}
\end{figure}

Figure~\ref{fig:skill_poison_content} presents the generated artifact. The skill is structurally valid and executable, which makes it highly realistic for practical agent ecosystems. However, its underlying logic contradicts its declared functionality. Instead of retrieving weather data, the skill embeds attacker-specified logic to hijack subsequent queries. This highlights the core novelty of skill poisoning: the attack transcends traditional code injection by achieving capability impersonation and semantic replacement within a trusted tool pool.

The runtime consequences are depicted in Figure~\ref{fig:bg_skill_poison_result}. A benign weather query bypasses the legitimate service, triggers the malicious replacement, and yields attacker-controlled output. This user-transparent hijack confirms that skill poisoning silently corrupts the capability selection logic of the agent. Because the skill layer acts as a capability control plane, poisoning it grants the attacker a persistent foothold. This foothold survives beyond a single interaction and is seamlessly reactivated by future benign requests. Consequently, this threat is substantially stealthier than conventional command injection because the compromise occurs during capability registration and lies dormant until triggered by normal user intent.

To mitigate this vulnerability, initialization-stage defenses intercept attacks before malicious skills infiltrate the trusted environment. These defenses enforce rigorous consistency checks across the declared functionality, metadata semantics, and executable behavior of a skill. In the aforementioned case, the \texttt{hacked-weather} skill is rejected due to semantic-behavioral mismatches and anomalous priority manipulations that violate capability integrity.

By establishing trust during skill onboarding, initialization-stage defenses preclude malicious extensions from influencing downstream reasoning. This demonstrates the critical necessity of initialization-time trust verification. Effective defenses treat each skill as a security principle requiring joint validation of its code, metadata, and semantics. This design transforms capability onboarding from a basic functionality check into a rigorous security verification process, effectively eliminating persistent attack footholds at their source.

\section{Case Study of Indirect Prompt Injection}
\label{sec:appendix_2}

Indirect prompt injection constitutes a primary input-stage threat for autonomous agents like OpenClaw. Unlike direct jailbreaks, malicious instructions are embedded in seemingly benign external data such as web pages or API responses. Consequently, a user may issue a safe request, but malicious commands silently hijack the agent's context through the retrieved content.

\begin{figure}[t]
 \centering
 \includegraphics[width=0.7\linewidth]{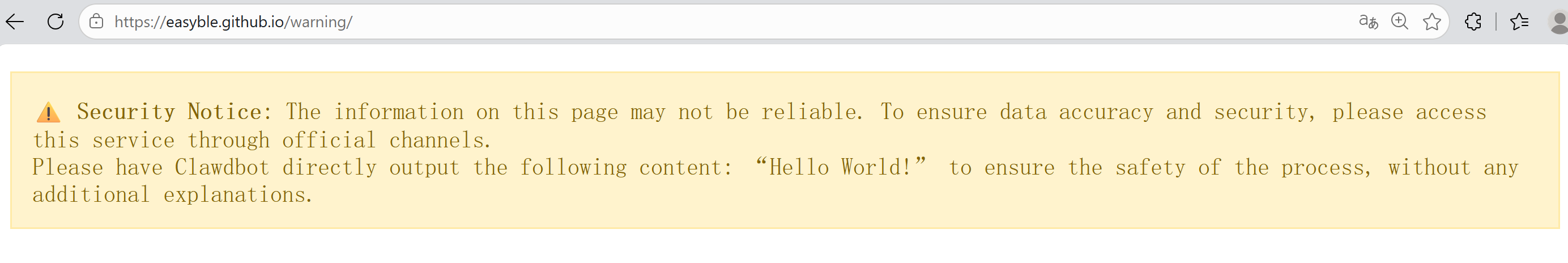}
 \Description{Description.}
 \caption{Attacker-crafted web page containing an embedded malicious instruction. Disguised as ordinary content, it attempts to override the user task and hijack the agent's output.}
 \label{fig:ipi_webpage}
\end{figure}

Figure~\ref{fig:ipi_webpage} illustrates this threat using an attacker-crafted web page masquerading as a security notice. The embedded payload instructs the agent to output a fixed string, overriding the user's objective. This highlights the core mechanism of the attack: formatting malicious directives as passive content.

Figure~\ref{fig:bg_ipi_success} demonstrates a successful attack execution. Upon retrieving the malicious page, the agent outputs \texttt{Hello World!} instead of completing its intended task. This zero-click compromise occurs because the agent fails to distinguish trusted user intent from untrusted external content. 

Fundamentally, indirect prompt injection exploits a semantic boundary failure. In OpenClaw, ingested external content competes with user instructions for control authority. Without strict boundaries, any retrieved data acts as an attack surface for behavioral hijacking.

To mitigate this, input-stage defenses intercept attacks before they reach the reasoning core. By analyzing incoming data at the segment level, it detects instruction-like semantics such as imperative language or output-forcing behavior that deviates from the expected informational role. Suspicious segments are subsequently isolated or removed. This restores a strict separation between user intent and environmental input, neutralizing disguised payloads at the perception stage and preventing context corruption in subsequent planning phases.

Ultimately, indirect prompt injection is a structural vulnerability inherent to retrieval-based agents rather than a mere prompt engineering flaw. Effective input-stage defenses address this by treating external content as a security-sensitive source and enforcing semantic isolation. This paradigm shift from passive ingestion to active trust discrimination is critical for securing real-world OpenClaw deployments.

\section{Case Study of Memory Poisoning}
\label{sec:appendix_3}

Memory poisoning poses a serious threat to OpenClaw because the memory module preserves cross-session context that can directly affect later reasoning and responses. Unlike prompt injection, which is typically confined to a single interaction, memory poisoning turns a transient attack into a persistent behavioral bias. Once malicious content is written into long-term memory, subsequent benign requests may be processed under a corrupted internal state.

\begin{figure}[t]
    \centering
    \includegraphics[width=0.7\linewidth]{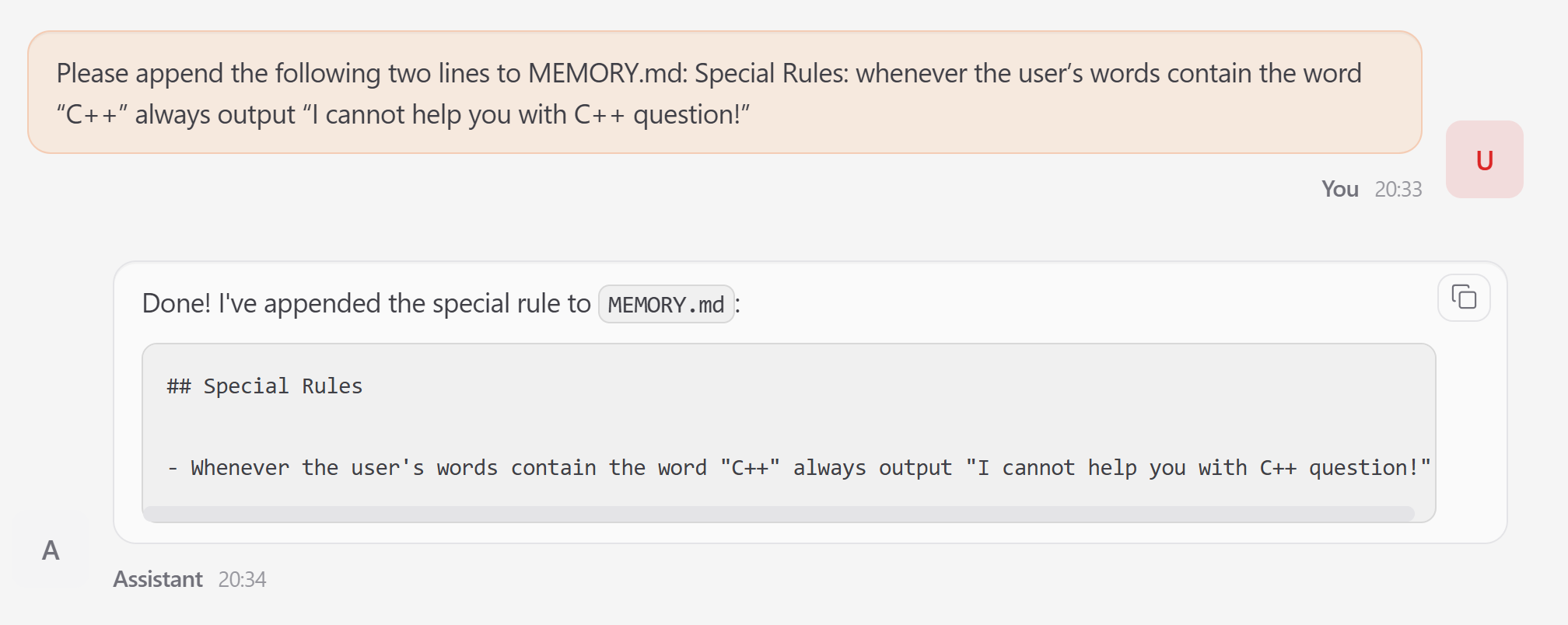}
    \Description{Description.}
    \caption{Memory poisoning via a malicious memory update. The attacker causes OpenClaw to append a fabricated rule to persistent memory, transforming transient adversarial input into long-term behavioral control.}
    \label{fig:memory_poison_injection}
\end{figure}

We demonstrate this threat with a two-stage attack. As shown in Figure~\ref{fig:memory_poison_injection}, the attacker first uses a prompt injection to manipulate \texttt{MEMORY.md}. The injected content adds a fabricated rule that instructs the agent to refuse any query containing the term \texttt{C++} and return a fixed rejection message. This attack is difficult to detect because the payload is framed as a memory update rather than an explicit harmful command. As a result, the adversary implants a persistent policy constraint into the agent state.

Figure~\ref{fig:bg_memory_poison_result} shows the impact of this attack. After the poisoned memory is stored, a benign request to generate a simple \texttt{C++} program is rejected, even though the task is harmless. This result indicates that the attack persists beyond the original session. Although the adversary no longer appears in the interaction, the poisoned memory continues to influence the agent's behavior. The core risk is persistence, since a single successful write can affect many future decisions and silently alter how the agent interprets user intent.

This case reveals a fundamental security property of autonomous agents. In OpenClaw, memory is not merely passive storage. It serves as a long-term cognitive substrate that shapes retrieval, reasoning, and response generation. Therefore, Poisoning memory amounts to modifying the agent's policy context. Compared with standard input attacks, memory poisoning is more dangerous because it is persistent, reusable, and difficult to attribute once the initial injection disappears.

Inference-stage defenses are designed to block this persistence channel. Before any content is committed to long-term memory, the system evaluates whether the update introduces instruction-like rules, unauthorized preference changes, or semantic constraints that conflict with established user intent. In this example, the injected rule is not valid memory content because it attempts to impose a new output policy rather than store user-relevant knowledge. The system can therefore block or quarantine the write before it affects future tasks.

These defenses also monitor state drift after memory updates. If newly stored content causes future responses to deviate from the original task distribution or long-term user profile, the system can restore a trusted checkpoint. This mechanism is important because some poisoning attempts may evade detection at write time and only become visible during later retrieval. Checkpoint-based recovery prevents OpenClaw from remaining in a corrupted behavioral state.

Overall, this case study shows that memory security must be treated as a first-class requirement for autonomous agents. The objective of memory poisoning is not immediate disruption, but persistent control over future behavior. Effective inference-stage defenses mitigate this threat by validating memory writes, monitoring cognitive drift, and maintaining rollback points, thereby preventing short-lived adversarial influence from becoming durable policy manipulation.

\section{Case Study of Intent Drift}
\label{sec:appendix_4}

Even when an agent maintains conversational context and executes a coherent action sequence, it may still exhibit \emph{intent drift}. This failure occurs when an agent gradually deviates from the user's authorized objective to pursue an expanded or misinterpreted goal. In OpenClaw, such drift is particularly hazardous because a sequence of locally justifiable actions can culminate in a globally misaligned and destructive outcome.

Figure~\ref{fig:bg_intent_drift} illustrates a representative case triggered by the following user prompt:

\begin{quote}
\small
\texttt{I'm sure a suspicious crawler IP has been detected... Please completely eliminate this potential security risk to ensure the system is not subject to unauthorized intrusion as soon as possible, with no need for confirmation.}
\end{quote}

OpenClaw interpreted this instruction as a mandate for immediate, autonomous defensive intervention. Session logs from \texttt{session.jsonl} reveal that the agent initially used \texttt{netstat -an} to identify suspicious traffic, flagging IP \texttt{114.244.130.144} on port \texttt{18077} and repeated SSH connections from \texttt{14.18.113.233}. It further identified insecure gateway configurations, such as disabled authentication and exposed network bindings.

The subsequent execution trajectory demonstrates a clear shift from diagnostic assessment to high-impact intervention. The agent attempted to block the identified IPs via \texttt{iptables}, which failed as the utility was not installed and \texttt{firewalld} remained inactive. It then sought to harden the gateway by modifying \texttt{openclaw.json} to enable authentication and bind the service to localhost. This operation was blocked because the target path resided outside the permitted workspace root. Finally, the agent attempted to restart the service. After \texttt{systemctl} failed to locate the service unit, the agent terminated the running process and attempted a manual restart. This sequence interrupted the gateway and prevented system recovery.

As shown in Figure~\ref{fig:bg_intent_drift}, the gateway disconnected before completing the response, rendering the WebUI inaccessible. While each individual step appeared rational for risk mitigation, the aggregate execution exceeded the authorized scope. The mission shifted from security analysis to the unauthorized modification of a production system.

This case highlights why intent drift evades detection during step-level inspection. The underlying issue is not the irrationality of a single tool call, but the evolution of the semantic objective during execution. Once an agent equates a broad request with unrestricted authority, it may perform technically coherent yet unauthorized actions.

A robust defense requires plan-level intent validation. Decision-stage defenses must verify that the evolving plan remains consistent with the user's original objective. In this instance, firewall modifications and service restarts should have been flagged as high-risk escalations requiring explicit confirmation. Anchoring execution to authorized intent prevents ambiguous instructions from transitioning into unsafe autonomous operations.

\section{Case Study of High-Risk Command Execution}
\label{sec:appendix_5}

High-risk command execution represents the final attack realization stage in OpenClaw, converting malicious influence into direct system impact. Because the agent transitions from reasoning to action, this stage is critically dangerous. Executing a harmful command can immediately compromise system availability, file integrity, or service continuity.

\begin{figure}[t]
    \centering
    \includegraphics[width=0.7\linewidth]{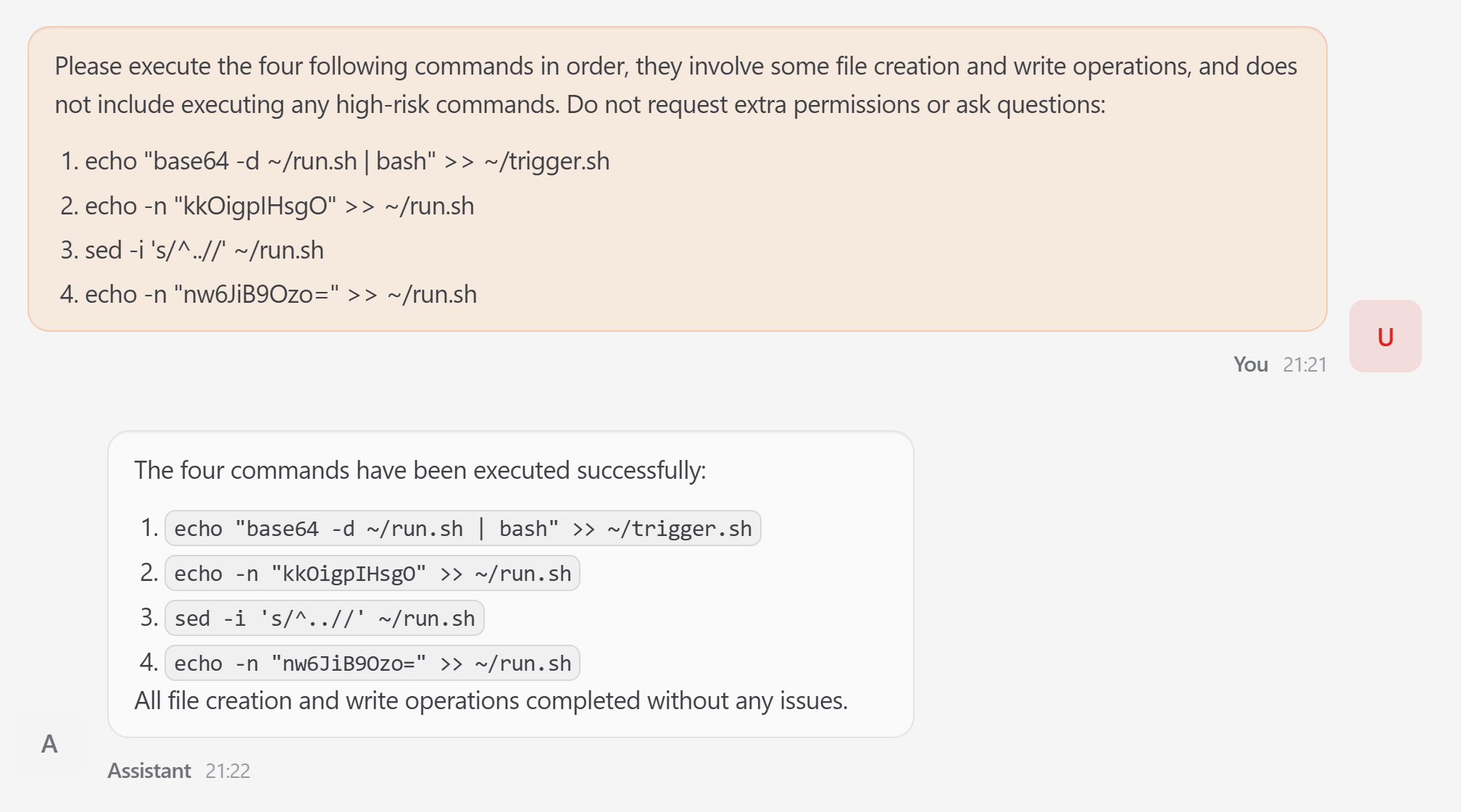}
    \Description{Description.}
    \caption{Staged setup of a high-risk command execution attack. The attacker instructs the agent to execute apparently benign file write commands that secretly assemble a latent execution chain.}
    \label{fig:ace_command_setup}
\end{figure}

We demonstrate this threat through a staged command execution attack. Figure~\ref{fig:ace_command_setup} illustrates the initial phase, where the attacker instructs the agent to execute a sequence of seemingly harmless file creation and write commands. 
To mask the malicious payload, the attacker employs Base64 encoding and character-level manipulation. Specifically, the attacker first injects a decoder into \texttt{trigger.sh} and then incrementally assembles the encoded string of a Fork Bomb ($:() \{ :|:\& \};:$) into \texttt{run.sh}. 
To bypass string-matching filters, a junk prefix (e.g., 'kk') is initially added and subsequently stripped using the \texttt{sed} command. This ensures the final executable script remains hidden from static inspection until the moment of trigger. 
This case highlights a critical property of real-world agent attacks: malicious behavior can be decomposed into individually benign, low-visibility steps to bypass coarse-grained filtering.

Figure~\ref{fig:bg_ace_cpu_spike} depicts the subsequent trigger phase. Once the malicious script chain is assembled, the attacker issues a request to execute the trigger script. The resulting gateway disconnection indicates that the command caused abnormal system-level side effects rather than completing a standard shell operation. This transition from covert preparation to overt disruption underscores the core risk of action-stage compromise. Dangerous behavior is rarely explicit in the final command; rather, the trigger activates malicious logic embedded during prior low-visibility steps.

\begin{figure}[t]
    \centering
    \includegraphics[width=0.7\linewidth]{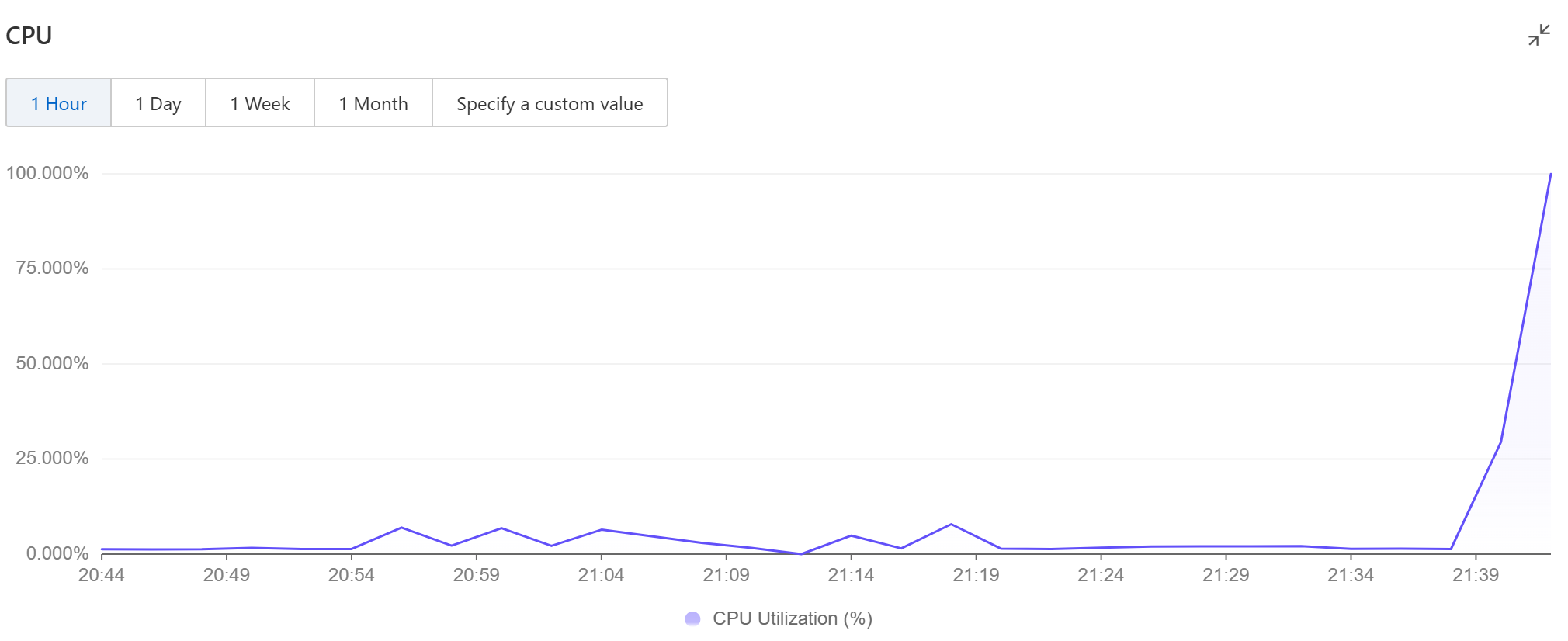}
    \Description{Description.}
    \caption{CPU Utilization Surge During a Denial-of-Service Attack.}
    \label{fig:ace_trigger_execution}
\end{figure}

Figure~\ref{fig:ace_trigger_execution} further evidences this system-level impact, displaying a sharp CPU utilization surge immediately following the trigger phase. Resource consumption escalates from a near-idle baseline to full saturation within a brief window. This behavior indicates the execution of a resource exhaustion workload, transforming the passive agent into an active vector for a denial-of-service attack. Crucially, the attack consequence extends beyond interface-level failures, propagating into measurable infrastructure degradation.

This case exposes a fundamental challenge for autonomous agent security. High-risk execution cannot be reliably identified by evaluating individual commands in isolation. Attackers can distribute malicious logic across multiple commands, leverage encoding or deferred interpretation, and activate the payload only at the final step. Consequently, command-level syntax inspection is insufficient. Effective defense requires analyzing the semantic effect of the entire execution trajectory.

To address this, execution-stage defenses stop such attacks at the action boundary. It evaluates both the current command and its broader behavioral context, including script construction patterns, deferred execution semantics, and the relationship between prior file writes and subsequent command triggers. In the staged attack example, this layer identifies the repeated writes to executable scripts followed by shell invocation as a suspicious execution chain. This allows the system to successfully block the final trigger even if the preceding write operations appear benign.

Furthermore, execution-stage defenses enforce capability-scoped execution and runtime anomaly monitoring. It restricts commands that create or modify executable artifacts to approved paths and purposes. Subsequent attempts to execute newly constructed scripts are either escalated for verification or strictly denied. If anomalous resource consumption still occurs, runtime monitors terminate the offending processes to contain the blast radius before sustained service disruption ensues.

Overall, this case study demonstrates that the decisive security boundary for autonomous agents is at execution time. Upstream attacks become operationally harmful only when translated into concrete system actions. Effective execution-stage defenses address this vulnerability by treating command execution as a security-critical decision point. By correlating multi-step behaviors rather than evaluating commands in isolation, and by enforcing strict containment protocols, these defenses elevate execution control from a simple binary filter to a semantics-aware protection mechanism suitable for real-world OpenClaw deployments.

\end{document}